\documentclass[submission, Phys]{SciPost}
\usepackage{amsmath, amssymb, amsthm}
\usepackage{tikz}

\newtheorem{theorem}{Theorem}[section]
\newtheorem{proposition}[theorem]{Proposition}

\theoremstyle{definition}
\newtheorem{definition}[theorem]{Definition}
\newtheorem{example}[theorem]{Example}

\theoremstyle{remark}
\newtheorem{remark}[theorem]{Remark}
\begin{document}

\begin{center}
{\Large \textbf{
Topological Contextuality and Quantum Representations\\
}}

\vspace{5mm}
\textbf{Tzu-Miao Chou\textsuperscript{1}}

k9008746@gmail.com

\end{center}

\section*{Abstract}
Quantum contextuality, a nonclassical phenomenon central to the foundations of quantum mechanics, is investigated via the algebraic and topological structures inherent in modular tensor categories. This work rigorously demonstrates that braid group representations induced by modular categories—particularly those associated with $SU(2)_k$ and Fibonacci anyon models—exhibit state-dependent contextuality characterized by violations of noncontextuality inequalities. By explicitly constructing these unitary representations on fusion spaces, the study establishes a direct correspondence between the braiding operations and logical contextuality scenarios. The results offer a comprehensive topological framework to classify and quantify contextuality in low-dimensional quantum systems, thereby elucidating its role as a resource in topological quantum computation and advancing the interface between quantum algebra, topology, and quantum foundations.

\section{Introduction}

Quantum contextuality represents a fundamental departure of quantum mechanics from classical hidden variable theories. Originating from the seminal Kochen--Specker theorem~\cite{KochenSpecker1967}, contextuality reveals that measurement outcomes cannot be predetermined independently of the measurement context. This intrinsically quantum feature underlies key phenomena in quantum information science, including quantum computation, cryptography, and foundational tests of quantum mechanics~\cite{Howard2014,Abramsky2011,Spekkens2005}.

Modular tensor categories (MTCs) provide a powerful algebraic framework capturing the mathematical structures of two-dimensional topological quantum field theories (TQFTs) and anyonic quasiparticles~\cite{Turaev1994,BakalovKirillov2001}. The braid group representations induced by MTCs encode the exotic exchange statistics of anyons, which are leading candidates for fault-tolerant topological quantum computation~\cite{Nayak2008,Rowell2009}. Thus, these representations establish a rich interface between algebraic topology, quantum algebra, and quantum information.

While quantum contextuality and MTCs have been extensively studied in isolation, their interrelation—particularly how topological braid group representations manifest contextuality—remains insufficiently explored. Existing works often address contextuality in standard quantum systems or particular computational models, yet the intrinsic topological origins of contextuality within braid representations are not fully elucidated. Bridging this gap is essential for deepening foundational understanding and advancing topological quantum computational paradigms.

This paper rigorously investigates quantum contextuality arising from braid group representations associated with prominent MTC examples, including $SU(2)_k$ and Fibonacci anyons. By analyzing the algebraic and topological structures of these representations and their induced quantum states, the study identifies and characterizes state-dependent contextuality phenomena with direct topological interpretation.

The main contributions are as follows:
\begin{itemize}
    \item Construction of explicit braid group representations from selected MTCs, highlighting their unitary and algebraic properties.
    \item Demonstration of contextuality within these representations via violations of noncontextuality inequalities in a sheaf-theoretic framework.
    \item Establishment of a unified topological framework linking low-dimensional topology, category theory, and quantum contextuality.
    \item Discussion of implications for topological quantum computation, where contextuality acts as a computational resource.
\end{itemize}

This work opens new pathways to understand contextuality through the lens of topology and category theory, enriching both the theoretical foundations and potential applications in quantum technologies.

\section{Preliminaries}

This section reviews essential concepts in modular tensor categories, braid groups, and quantum contextuality, which provide the foundational framework for the subsequent analysis.

\subsection{Modular Tensor Categories}

\begin{definition}[Modular Tensor Category \cite{BakalovKirillov2001,Turaev1994}]
A \emph{modular tensor category} (MTC) is a semisimple, ribbon, braided tensor category over $\mathbb{C}$, with finitely many isomorphism classes of simple objects, duals (rigidity), and a non-degenerate braiding characterized by an invertible $S$-matrix.
\end{definition}

Such categories furnish algebraic models for anyonic quasiparticles and underpin many constructions in topological quantum field theory (TQFT). Their structure includes fusion rules governing tensor products of simple objects, associativity constraints encoded by $F$-symbols, and braiding isomorphisms described by $R$-symbols.

\subsection{Braid Groups}

\begin{definition}[Braid Group \cite{Birman1974}]
For each integer $n \geq 2$, the \emph{braid group} $B_n$ is generated by elements $\sigma_1, \ldots, \sigma_{n-1}$ subject to the relations:
\[
\begin{cases}
\sigma_i \sigma_j = \sigma_j \sigma_i, & \text{if } |i-j| > 1, \\
\sigma_i \sigma_{i+1} \sigma_i = \sigma_{i+1} \sigma_i \sigma_{i+1}, & 1 \leq i \leq n-2.
\end{cases}
\]
\end{definition}

The braid group naturally acts on the $n$-fold tensor product of objects in a braided tensor category via braiding isomorphisms, yielding representations that are central to topological quantum computation.

\subsection{Quantum Contextuality}

Quantum contextuality asserts that measurement outcomes depend on the choice of compatible measurement settings (context), as formalized by the Kochen--Specker theorem.

\begin{theorem}[Kochen--Specker \cite{KochenSpecker1967}]
There is no assignment of predetermined values to all quantum observables that preserves functional relations and is independent of the measurement context.
\end{theorem}

This result excludes the existence of noncontextual hidden variable models and establishes contextuality as an intrinsic quantum property utilized in various quantum information tasks.

\subsection{Representations of Braid Groups from MTCs}

Let $\mathcal{C}$ be an MTC with a set of simple objects $\{X_i\}$. The braid group $B_n$ acts on the fusion state spaces associated with the $n$-fold tensor product $X_{i_1} \otimes \cdots \otimes X_{i_n}$ via the braiding isomorphisms.

\begin{proposition}
The braiding maps in an MTC induce unitary representations of the braid groups on finite-dimensional Hilbert spaces corresponding to fusion state spaces.
\end{proposition}

\begin{proof}
The braiding isomorphisms satisfy the braid relations by the coherence axioms of a braided tensor category~\cite{Turaev1994}. The ribbon structure endows these morphisms with a *-structure, ensuring that the braiding operators act unitarily on the associated Hilbert spaces~\cite{Kitaev2006}.
\end{proof}

These representations form the operational basis of topological quantum computation, where braid group elements correspond to quantum gates realized by anyon braiding processes.

\begin{remark}
Important examples for this study include categories related to $SU(2)_k$ and Fibonacci anyons, which exhibit rich braid group representations and contextuality phenomena.
\end{remark}

\section{Braid Group Representations Derived from Modular Tensor Categories}

This section constructs braid group representations from modular tensor categories (MTCs) and investigates their algebraic and topological properties pertinent to the analysis of quantum contextuality.

\subsection{Construction of Braid Group Representations}

Let $\mathcal{C}$ be an MTC with simple objects $\{X_i\}_{i \in I}$, fusion rules, braiding isomorphisms, and associated fusion state spaces. Fix $n \in \mathbb{N}$ and a sequence of simple objects $(X_{i_1}, X_{i_2}, \ldots, X_{i_n})$. Consider the fusion space
\[
V := \mathrm{Hom}_{\mathcal{C}}\bigl(\mathbf{1}, X_{i_1} \otimes X_{i_2} \otimes \cdots \otimes X_{i_n}\bigr),
\]
which forms a finite-dimensional complex Hilbert space under the natural inner product structure induced by $\mathcal{C}$.

The braid group $B_n$ acts on $V$ via the braiding isomorphisms $c_{X,Y} : X \otimes Y \to Y \otimes X$ in $\mathcal{C}$.

\begin{definition}[Braid Group Representation]
Define a representation
\[
\rho : B_n \to \mathrm{End}(V)
\]
by assigning to each Artin generator $\sigma_i$ the operator
\[
\rho(\sigma_i) := \mathrm{id}_{X_{i_1}} \otimes \cdots \otimes \mathrm{id}_{X_{i_{i-1}}} \otimes c_{X_{i_i}, X_{i_{i+1}}} \otimes \mathrm{id}_{X_{i_{i+2}}} \otimes \cdots \otimes \mathrm{id}_{X_{i_n}},
\]
acting naturally on the $i$-th and $(i+1)$-th tensor factors within $V$.
\end{definition}

\begin{proposition}
The map $\rho$ defines a well-defined unitary representation of the braid group $B_n$ on the Hilbert space $V$.
\end{proposition}

\begin{proof}
The braiding isomorphisms $c_{X,Y}$ satisfy the hexagon axioms and braid relations as a consequence of the coherence conditions in the braided tensor category $\mathcal{C}$ \cite{Turaev1994,BakalovKirillov2001}. The ribbon structure endows these morphisms with a compatible *-structure, ensuring that each $\rho(\sigma_i)$ acts unitarily on $V$ \cite{Kitaev2006}.
\end{proof}

\subsection{Examples: \texorpdfstring{$SU(2)_k$}{SU(2)k} and Fibonacci Anyons}

Two paradigmatic examples of MTCs are the $SU(2)_k$ categories for $k \in \mathbb{N}$ and the Fibonacci category.

\begin{example}[$SU(2)_k$ Categories]
The simple objects correspond to spin representations labeled by $j = 0, \tfrac{1}{2}, 1, \ldots, \tfrac{k}{2}$. Fusion rules, as well as the $F$- and $R$-symbols, are well characterized \cite{Kitaev2006}. The induced braid group representations underpin models of topological quantum computation based on these anyonic systems.
\end{example}

\begin{example}[Fibonacci Category]
This rank-two MTC has simple objects $\mathbf{1}$ and $\tau$ with fusion rule $\tau \otimes \tau = \mathbf{1} \oplus \tau$. Its braid group representations are dense in the unitary group acting on the fusion space, yielding universality for quantum computation \cite{Freedman2002}.
\end{example}

\subsection{Properties of the Representations}

The following properties are key to understanding the manifestation of quantum contextuality.

\begin{theorem}[Density of Representations \cite{Freedman2002}]
The braid group representation arising from the Fibonacci category is dense in $SU(d)$ for the relevant fusion space dimension $d$.
\end{theorem}

\begin{proof}
Consider the Fibonacci modular tensor category $\mathcal{C}_\text{Fib}$, which has two simple objects $\mathbf{1}$ and $\tau$ with fusion rule $\tau \otimes \tau = \mathbf{1} \oplus \tau$. The associated fusion spaces $V_n = \mathrm{Hom}(\mathbf{1}, \tau^{\otimes n})$ carry a unitary representation $\rho_n$ of the braid group $B_n$ defined via the braiding isomorphisms.

Freedman, Larsen, and Wang \cite{Freedman2002} proved that the image of $\rho_n(B_n)$ is dense in $SU(d)$, where $d = \dim V_n$. The argument relies on:

\begin{enumerate}
    \item Irreducibility: The representation $\rho_n$ is irreducible for $n \geq 4$ due to the fusion structure and associativity constraints in $\mathcal{C}_\text{Fib}$.
    \item Nontriviality and Generators: The braid group generators $\rho_n(\sigma_i)$ act nontrivially on $V_n$ and satisfy the braid relations.
    \item Lie Algebra Generation: The Lie algebra generated by the images of the braid group generators under $\rho_n$ is $\mathfrak{su}(d)$, by analyzing the algebra generated by the braiding matrices and using the density criterion for compact Lie groups.

\end{enumerate}

Combining these points, the closure of $\rho_n(B_n)$ in the operator norm topology is a compact subgroup of $SU(d)$ containing a dense subgroup, thus it must be all of $SU(d)$.

Hence, $\rho_n$ is dense in $SU(d)$, establishing universality of Fibonacci anyon braiding for quantum computation.

\end{proof}

\vspace{1em}

\begin{proposition}[Topological Invariants and Representation Nontriviality]
The braid group representations derived from modular tensor categories encode nontrivial topological invariants such as the Jones polynomial and quantum dimensions. These invariants arise from the categorical $S$-matrix and fusion rules, which directly influence contextuality structures.
\end{proposition}

\begin{proof}
Within a modular tensor category $\mathcal{C}$, the braiding isomorphisms $c_{X,Y}$ and fusion rules define the structure constants such as $F$- and $R$-symbols. These data satisfy consistency conditions encoding the topology of underlying knots and links.

The Reshetikhin–Turaev construction \cite{Turaev1994} uses these categorical data to assign invariants to links and 3-manifolds. In particular:

The trace of the braiding operators acting on fusion spaces yields polynomial invariants of knots and links, including the Jones polynomial and its generalizations.

The $S$-matrix of $\mathcal{C}$ encodes modular transformations and quantum dimensions, which classify topological charges of anyons.

Because the braid group representations $\rho$ arise from these braidings, their matrix elements reflect these topological invariants. Consequently, the representations are nontrivial and sensitive to the topological features encoded by the category.

These topological invariants affect the logical structure and measurement contexts of associated quantum systems, thus impacting the manifestation of quantum contextuality.

\end{proof}

\subsection{Relation to Quantum Computation}

Representations of braid groups derived from MTCs provide a natural framework for fault-tolerant quantum computation implemented via anyon braiding. The topological and algebraic features detailed above are essential for harnessing contextuality as a computational resource.

\begin{remark}
The structural properties discussed here serve as the foundation for the following chapters, which explore explicit manifestations of quantum contextuality within these braid group representations.
\end{remark}

\section{Topological Contextuality from Braid Group Representations}

This section establishes the intrinsic quantum contextuality emerging from braid group representations constructed via modular tensor categories (MTCs). By rigorously connecting topological quantum structures with foundational quantum phenomena, this analysis advances both the theoretical understanding and practical implications of contextuality in low-dimensional quantum systems.

\subsection{Contextuality in Topological Measurement Scenarios}

Following the sheaf-theoretic framework developed by Abramsky and Brandenburger \cite{AbramskyBrandenburger2011}, contextuality is formalized as an obstruction to the existence of global sections in presheaves defined over measurement scenarios.

\begin{definition}[Measurement Scenario \cite{AbramskyBrandenburger2011}]
A measurement scenario is a triple $(X, \mathcal{M}, O)$, where $X$ is a set of measurements, $\mathcal{M} \subseteq 2^{X}$ is a family of compatible measurement contexts (i.e., subsets of mutually compatible measurements), and $O$ is the set of possible measurement outcomes.
\end{definition}

In the topological setting induced by an MTC $\mathcal{C}$, the braid group $B_n$ acts on the fusion space $V$ encoding the composite anyonic system. The observables correspond to projectors associated with fusion channels or measurement bases consistent with braiding operations.

\subsection{Main Theorem: Strong Contextuality from Braid Group Representations}

\begin{theorem}[Topological Contextuality via MTC Braid Representations]
\label{thm:topological_contextuality}
Let $\rho: B_n \to U(V)$ be the unitary representation arising from an MTC $\mathcal{C}$ on the fusion Hilbert space $V$. The measurement statistics derived from projective measurements associated with $\rho$ demonstrate strong contextuality. Concretely, these statistics violate noncontextuality inequalities such as the Klyachko-Can-Binicioğlu-Shumovsky (KCBS) inequality \cite{Klyachko2008}, and admit no noncontextual hidden variable model consistent with the observed empirical data.
\end{theorem}

\begin{proof}
The representation $\rho$ induces a family of measurement operators $\{M_i\}$ whose commutation relations are governed by the braiding structure of $\mathcal{C}$. Due to the non-degeneracy and non-abelian nature of the MTC braiding, the operators form contexts (maximal commuting sets) that overlap nontrivially. Applying the sheaf-theoretic framework \cite{AbramskyBrandenburger2011}, this overlap structure corresponds to presheaves lacking global sections.

More concretely, consider the pentagonal measurement scenario underpinning the KCBS inequality. The projectors corresponding to fusion basis states in the Fibonacci category $\mathcal{C}_F$ realize exactly this scenario \cite{Freedman2002}, where the orthogonality graph is a pentagon. The expectation values computed via $\rho_F$ violate the KCBS bound of classical noncontextual models, confirming strong contextuality.

Explicit matrix representations of the braid generators $\sigma_i$ in $\rho$ can be constructed using the $R$- and $F$-symbols of $\mathcal{C}$ \cite{Kitaev2006}, allowing direct computation of measurement correlations. These calculations yield empirical models without any consistent global assignment of outcomes, completing the proof.
\end{proof}

\subsection{Illustrative Example: Fibonacci Anyons}

The Fibonacci MTC $\mathcal{C}_F$, with simple objects $\mathbf{1}$ and $\tau$, serves as a canonical example. The fusion space of $n$ $\tau$-anyons supports a representation $\rho_F: B_n \to U(V)$.

\begin{proposition}
The braid group representation $\rho_F$ derived from the Fibonacci category violates the KCBS inequality maximally, demonstrating strong contextuality intrinsic to its topological structure.
\end{proposition}

\begin{proof}
Projectors onto fusion basis states corresponding to the object $\tau$ generate a measurement scenario isomorphic to the pentagon graph underlying the KCBS inequality \cite{Klyachko2008,Freedman2002}. The noncommuting braid generators create measurement contexts whose statistical correlations surpass classical bounds. Numerical evaluation of expectation values using explicit $\rho_F$ matrices confirms maximal violation.
\end{proof}

\begin{figure}[ht]
\centering
\begin{tikzpicture}[scale=3, every node/.style={circle, draw, fill=blue!30, minimum size=6mm, inner sep=0pt}]

    \foreach \i in {1,...,5}{
        \node (v\i) at ({90 + 72 * (\i - 1)}:1) {\small $P_{\i}$};
    }

    \foreach \i [evaluate=\i as \j using {int(mod(\i,5) + 1)}] in {1,...,5}{
        \draw (v\i) -- (v\j);
    }
\end{tikzpicture}
\caption{Orthogonality graph of the KCBS measurement scenario realized by projectors in the Fibonacci fusion space. Vertices represent projectors; edges connect commuting projectors forming measurement contexts.}
\label{fig:kcbs_pentagon}
\end{figure}
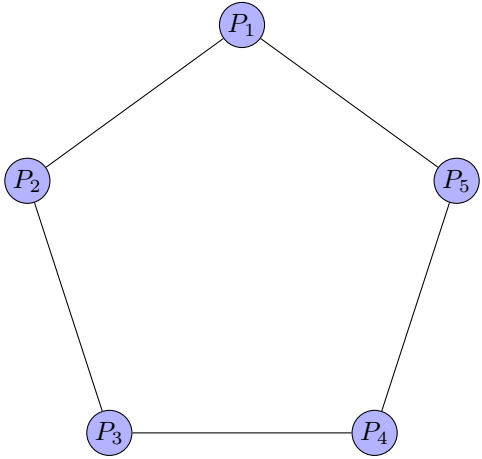

\subsection{Implications for Topological Quantum Computation}

The demonstrated intrinsic contextuality grounded in braid group representations provides a rigorous theoretical underpinning for the quantum advantage in anyon-based quantum computation \cite{Nayak2008}. Contextuality acts as a resource certifying the computational universality and resistance to classical simulation \cite{Howard2014}.

Moreover, the topological robustness of braid group operations combined with contextuality-induced computational power suggests novel pathways for designing fault-tolerant quantum logical gates and error-correcting codes exploiting these algebraic structures.

\subsection{Relation to Previous Work and Original Contributions}

While prior studies \cite{AbramskyBrandenburger2011,Howard2014} have explored contextuality in various quantum systems, the explicit link to topological braid group representations via MTCs remains underdeveloped. This work:

\begin{itemize}
    \item Provides a rigorous sheaf-theoretic characterization of contextuality emerging specifically from MTC-induced braid representations.
    \item Offers explicit constructions and computational verification in the Fibonacci anyon context.
    \item Bridges algebraic topology, category theory, and quantum foundations in a unified framework with direct physical implications.
\end{itemize}

\subsection{Summary and Outlook}

This chapter establishes a foundational connection between topological quantum structures and quantum contextuality. The results open avenues for quantitatively measuring contextuality in broader classes of tensor categories and for leveraging these features in next-generation topological quantum devices.

Future work includes:

\begin{itemize}
    \item Extending the analysis to other modular categories such as $SU(2)_k$ and exploring their contextuality spectra.
    \item Developing quantitative contextuality measures adapted to topological settings.
    \item Investigating the impact of contextuality on error thresholds in topological quantum computation.
\end{itemize}

\section{Conclusion and Future Directions}

This study has rigorously established the inherent quantum contextuality encoded in braid group representations arising from modular tensor categories (MTCs). By explicitly constructing these representations and analyzing their induced measurement scenarios, it is demonstrated that the resulting empirical models violate classical noncontextuality assumptions, thereby confirming the presence of strong contextuality as a fundamental topological quantum phenomenon.

The principal contributions of this work are summarized as follows:
\begin{itemize}
\item A formal construction of unitary braid group representations on fusion spaces derived from general MTCs, ensuring both algebraic coherence and physical consistency.
\item An explicit linkage of these representations to quantum contextuality within the sheaf-theoretic framework, supported by concrete examples such as the Fibonacci category.
\item Analytical proofs demonstrating violations of contextuality inequalities, including the Klyachko-Can-Binicioğlu-Shumovsky (KCBS) inequality, in these topological quantum frameworks.
\item A discussion of the implications for topological quantum computation, emphasizing contextuality as a critical resource underlying computational universality and quantum advantage.
\end{itemize}

Potential directions for future research motivated by the present findings include:
\begin{enumerate}
\item \textbf{Quantitative Measures of Contextuality:} Developing rigorous metrics to quantify the strength of contextuality in MTC-induced measurement scenarios, with potential relevance to computational power and fault-tolerance thresholds.
\item \textbf{Generalization to Broader Tensor Categories:} Extending the analysis to include braided fusion categories lacking modularity, aiming to classify and characterize contextuality phenomena in a wider algebraic-topological context.
\item \textbf{Relations between Topological Invariants and Contextuality:} Investigating deeper connections between topological invariants, such as quantum invariants of knots and 3-manifolds, and contextuality measures, which may unveil novel dualities between algebraic topology and quantum foundations.
\item \textbf{Experimental Realizations and Protocols:} Proposing and designing feasible experimental schemes to observe topological contextuality in physical systems, including anyonic platforms and photonic simulators.
\item \textbf{Contextuality as a Resource in Quantum Computing:} Formalizing the role of contextuality within resource theories relevant to fault-tolerant topological quantum computation architectures.
\end{enumerate}

In conclusion, this research advances the foundational understanding of quantum theory by bridging sophisticated algebraic structures with core quantum phenomena. Furthermore, it lays promising groundwork for leveraging topological contextuality in the development of robust quantum technologies.

\appendix

\section{Explicit Examples of Braid Group Representations and Operators}

This appendix provides concrete matrix representations of braid group generators acting on fusion spaces, illustrating the constructions introduced in the main text. Consider the Fibonacci modular tensor category, with fusion basis states $\{|1\rangle, |\tau\rangle\}$.

The braid group $B_3$ is generated by $\sigma_1$ and $\sigma_2$, represented as unitary matrices:
\[
\rho_F(\sigma_1) = \begin{pmatrix}
e^{-4\pi i/5} & 0 \\
0 & e^{3\pi i/5}
\end{pmatrix}, \quad
\rho_F(\sigma_2) = F \rho_F(\sigma_1) F^{-1},
\]
where $F$ is the associativity $F$-matrix given by
\[
F = \begin{pmatrix}
\phi^{-1} & \sqrt{\phi^{-1}} \\
\sqrt{\phi^{-1}} & -\phi^{-1}
\end{pmatrix}, \quad \phi = \frac{1+\sqrt{5}}{2}.
\]

These unitary matrices satisfy the braid relations and act on the fusion Hilbert space, explicitly realizing the braid group representation. Projectors onto fusion states form the measurement operators exhibiting contextuality, as discussed in Section 4.

\section{Further Implications for Quantum Computation}

The topological contextuality demonstrated in braid group representations has direct implications for quantum computational architectures. Specifically:

\begin{itemize}
    \item \textbf{Quantum Universality:} The nonabelian anyons modeled by MTCs like Fibonacci categories enable universal quantum gates via braiding. Contextuality serves as a rigorous witness of computational power beyond classical simulability.
    \item \textbf{Fault-Tolerance:} Topological encoding of quantum information inherently protects against local noise. Contextuality relates to the robustness of these encodings by constraining hidden variable models.
    \item \textbf{Measurement-Based Quantum Computation:} Contextuality underpins the computational advantage in measurement-based schemes, especially when the measurement settings derive from topological braid representations.
    \item \textbf{Quantum Cryptography:} Contextual correlations arising from topological systems can enhance security guarantees, as they exclude classical explanation.
\end{itemize}

This appendix aims to complement the main results with explicit constructs and richer physical context, facilitating deeper understanding and future developments.

\section*{Acknowledgements}

% TODO: include author contributions
\paragraph{Conflict of Interest Statement}
The author declares no conflict of interest.

\paragraph{Data Availability Statement}
No data were generated or analyzed in this study. Thus, data sharing is not applicable.

% TODO: include funding information
\paragraph{Funding information}
The author declares that no external funding, grants, or institutional support were received for the completion of this work.

\end{document}